\documentclass[12pt,preprint]{aastex}

\slugcomment{}

\shorttitle{Multi-wavelength Differential Imaging Experiment}
\shortauthors{Biller et al.}  

\begin{document}

\title{
A Multi-wavelength Differential Imaging Experiment for the High Contrast Imaging Testbed}

\author{Beth Biller$^{1,4}$, John Trauger$^2$, Dwight Moody$^2$, 
Laird Close$^3$, Andreas Kuhnert$^2$, Karl Stapelfeldt$^2$, 
Wesley A. Traub$^2$, Brian Kern$^2$}

\affil{$^1$ Institute for Astronomy, University of Hawaii at Manoa, Honolulu, HI 96822}
\affil{$^2$ Jet Propulsion Laboratory, California Institute of Technology}
\affil{$^3$ Steward Observatory, University of Arizona, Tucson, AZ 85721}
\affil{$^4$ Hubble Fellow}

\begin{abstract}
We discuss the results of a multiwavelength differential imaging 
lab experiment with the 
High Contrast Imaging Testbed (HCIT) at the Jet Propulsion Laboratory.  
The HCIT
combines a Lyot coronagraph with a Xinetics deformable
mirror in a vacuum environment  
to simulate a space telescope in order to test technologies and algorithms
for a future exoplanet coronagraph mission.  At present, ground
based telescopes have achieved 
significant attenuation of speckle noise using the 
technique of spectral differential imaging (SDI).
We test whether ground-based SDI can be generalized
to a non-simultaneous spectral 
differential imaging technique (NSDI) for a space mission.  In our 
lab experiment, a series of 5 
filter images centered around the 
O$_2$(A) absorption feature at 0.762 $\mu$m were
acquired at nominal contrast values of 
10$^{-6}$, 10$^{-7}$, 10$^{-8}$, and 10$^{-9}$.  
Outside the dark hole, single differences of images improve 
contrast by a factor of $\sim$6.
Inside the dark hole, we
found significant speckle chromatism as a function of wavelength
offset from
the nulling wavelength, leading to a contrast 
degradation by a factor of 7.2 across the entire $\sim$80 nm
bandwidth.  This effect likely stems from 
the chromatic behavior of the current occulter.  New, 
less chromatic occulters are currently in development;
we expect that these new occulters will resolve the speckle chromatism 
issue.
\end{abstract}

\section{Introduction}

While over 300 exoplanets have now been discovered indirectly via
radial velocity, transits, or gravitational lensing, to date only a handful 
of planets have been directly imaged around a main sequence or pre-main sequence
star \citep{Kal08,Mar08}.  Most other detection methods (e.g. radial velocity and transit)
are more sensitive to close-in planets than those at greater distance
from their stars.  Direct imaging is the only method currently capable
of detecting planets $>$10 AU from their parent stars.  Numerous
adaptive optics and space based surveys to directly image young
Jupiter mass planets in the near-IR \citep{Bil07,Laf07,Far07}
have recently been completed; these
surveys have placed strong statistical limits on extrasolar planet
distributions but have produced no planet detections.  Clearly, directly imaging
planets is a nontrivial undertaking -- and one that becomes even 
more difficult for terrestrial planets.

In order to successfully image a planet one must first overcome:
1) the huge contrast difference between star and planet (the very 
youngest Jovian planets will be 10$^{-5}$ times fainter than their 
parent star, while a terrestrial planet will be $\sim$10$^{-10}$ times 
fainter). Modern approaches to planet-finding often employ a Lyot 
coronagraph to achieve these contrasts \citep[e.g.~][]{Kuc02}.  
2) Even once the initial star-planet contrast
has been achieved, residual superspeckles remain 
(from errors in the wavefront due to imperfect optics)
and produce a limiting ``speckle'' 
noise floor which prevents planet detection. 
 This is seen both from the ground and 
space (e.g. the ``breathing'' effect seen by HST).

A number of instrumental speckle suppression techniques have been 
suggested and implemented on ground base telescopes (and in 
testbed scenarios for space based telescopes), including
azimuthal differential imaging \citep[aka~ADI~or~roll~subtraction,~see~e.g.][]{Mar06,Laf07},   simultaneous spectral differential imaging  \citep[SDI,~see~e.g.][]{Mar05, Bil07, Len04, Chu08}, 
imaging polarimetry \citep{Per08, Tha08, Opp08}, 
various methods of focal plane wavefront sensing \citep{Guy06, Guy09, Cod08, Ken06}, 
and the use of Integrated Field Spectrographs (IFS) to separate speckles (which modulate
with wavelength) from real companions (which do not)  \citep{Lar06, McE07, McE08, Cla08}.
In this paper we focus on the possible application of the SDI technique 
to a space-based telescope platform.

From the ground, the SDI technique has been used after  
adaptive optics (AO) correction to remove residual speckle noise without 
removing planet light.  SDI has been implemented on the 
TRIDENT device at the CFHT \citep{Mar05}, on the
Simultaneous Differential Imager at the VLT and MMT \citep{Bil07,Len04} 
and, more recently, with a coronagraph,
on the Near-Infrared Coronagraphic Imager at Gemini South \citep{Chu08}.
SDI attenuates speckle noise by exploiting a spectral feature in the desired
target (previous implementations utilize the 1.6 $\mu$m methane absorption
feature robustly observed in substellar objects cooler than 1400
K). Images are taken simultaneously both within and outside the chosen
absorption feature. Due to the simultaneity of the observations, the
star and the coherent speckle pattern are largely identical in both
filters, while a faint companion with that absorption feature is
bright in the continuum filter and faint in the absorbed filter. 
Subtracting the absorbed filter image from the continuum filter image
thus removes the starlight and speckle patterns while a real companion
with the chosen absorption feature remains in the subtraction 
image. In other
words, the off-absorption band image acts as an ideal reference point
spread function (henceforth PSF) for the absorption band
image. Utilizing a signature spectral feature of substellar objects
greatly reduces the number of false positives detected (e.g. a
faint background object, while real, will drop out of the SDI subtraction
since it will not have methane absorption).  From the ground, 
significant attenuation of speckle noise has been achieved with SDI.
For example, \citet{Bil07} achieved H band contrasts $>$25000 (5
$\sigma$ $\Delta$F1(1.575 $\mu$m) $>$ 10 mag, $\Delta$H $>$ 10.8 mag for a T6
spectral type) at a separation of 0.5" from the primary star.  
With this degree of attenuation, it is possible to image (5 $\sigma$
detection) a 2 Jupiter mass planet at 5 AU around a 10 Myr M0 star at
10 pc \citep{Bil08, For08}.

SDI on ground based telescopes provides significant speckle
attenuations down to star-planet contrasts of $\sim$0.3-1$\times$10$^{-4}$.
To test the classical SDI technique at contrasts of 10$^{-6}$ to 10$^{-9}$ 
and to determine whether such a technique would be applicable for the
detection of terrestrial planets for a possible exoplanet coronagraph, 
e.g. a Terrestrial Planet
Finder mission, we implemented a similar multiwavelength differential
imaging scheme for the High Contrast Imaging Testbed 
\citep[henceforth~HCIT,~described~in~detail~in~the~supplemental~material~to][]{TT07}
at the Jet Propulsion Laboratory.  The HCIT
combines a band-limited Lyot coronagraph with a 32$\times$32 actuator
Xinetics deformable
mirror in a vacuum environment \citep{Tra04} 
to simulate a space telescope in order to test technologies and algorithms
for future coronagraph missions.  The HCIT has 
demonstrated extremely high contrasts and very fine wavefront control; 
contrasts of 10$^{-10}$ have been achieved, sufficient to image an
Earth-like extrasolar planet \citep{TT07}.  

\section{Experimental Design}

Five 2$\%$ bandwidth filters were selected near the
prominent O$_2$(A) absorption feature at 0.762 $\mu$m (seen in Earth's
atmosphere and expected for any terrestrial extrasolar planet with an
oxygen atmosphere, \citet{Woo02}) with approximate central wavelengths of
F1(768 nm), F2(784 nm), F3(800 nm), F4(816 nm), and F5(832 nm).
Exact filter wavelengths and bandwidths are presented in
Table~\ref{tab:filtprop} -- however, approximate wavelengths are used 
to refer to the filters throughout the text.  Two sets of images were taken in
each filter -- a long set, with 90 s exposure time per filter, and a
short set, with 5 s exposure time per filter. The filter set was
spaced across a considerable bandwidth in wavelength in order to
measure speckle chromatism as a function of wavelength.  The 
light source used for this experiment is a passively Q-switched 
Nd:YAG 1060 nm laser.  This laser delivers subnanosecond pulses into a single-mode 
supercontinuum photonic crystal fiber, producing light with a relatively smooth spectrum from 
600 nm to 1600 nm and providing sufficient bandwidth and stability to conduct this experiment \citep{TT07}.

For ground based observing with adaptive optics (AO),
quasi-static superspeckles due to residual instrumental 
errors remain even after AO correction \citep{Mar05,Mas05}.  These superspeckles 
are stable on timescales of minutes to hours. but 
still vary quickly enough over a typical observation
\citep[1-2~hours,~see~e.g.][]{Hin07, Mar06} that  a reference PSF must be built on similar
timescales in order to overcome the stochastic speckle noise floor.
(To build this PSF, ADI uses rotation on the sky to decorrelate speckles from real object,
SDI uses wavelength diversity, and imaging polarimetry uses images taken in different polarization states.) Thus, on the ground, simultaneous imaging in several bandwidths is desirable 
to successfully implement spectral differential imaging (SDI).
For space-based observing, however, 
speckles are stable on much longer 
timescales of hours to days, making simultaneity of 
imaging less necessary \citep[e.g.~][]{Spa02}.  While proposed instrument designs 
(such as those for the Terrestrial Planet Finder and the Eclipse mission)
utilize simultaneous imaging via either an integral field unit 
or dichroic beamslitter, no multispectral backend imager was 
available at the HCIT.   

\citet{TT07}  demonstrated extreme speckle
stability at the HCIT during a "movie experiment" where 480 
individual snapshot images were recorded over a 5 hour period.
They found that the change in speckle contrast in HCIT 
during this period (where the apparatus was allowed to drift
freely) was $\sim$0.1$\times$10$^{-10}$ per 5 hours.
This demonstrated speckle 
stability allowed the current experiment to be conducted non-simultaneously with 
valid results.  Additionally, the use of the same imaging channel 
avoids non-common
path wavefront errors between images and thus allows for a cleaner 
determination of inherent speckle chromaticity.
The present multi-wavelength 
differential imaging experiment measures speckle evolution as a function of
wavelength and contrast level.  We test whether 
the ground-based simultaneous spectral 
differential imaging technique can be generalized
to a non-simultaneous spectral 
differential imaging technique for a space mission. 

By using a 5 filter set, we can attempt to correct
for speckle chromatism using both single differences of images and 
the double difference technique pioneered by \citet{Mar00}.  
Since the radial position of the speckle pattern in 
each filter image is proportional to $\frac{\lambda}{D}$, the 
platescale of each image must be scaled so that the speckles in 
each filter fall at the same radii despite chromatic differences.
After this scaling, a number of single differences as well as 
a double difference of
images in the F1 through F5 filters are calculated:

\begin{equation}
I_{31}\equiv(F3 - F1) 
\end{equation}

\begin{equation}
I_{32}\equiv(F3 - F2) 
\end{equation}

\begin{equation}
I_{34}\equiv(F3 - F4)
\end{equation}

\begin{equation}
I_{35}\equiv(F3 - F5)
\end{equation}

\begin{equation}
I_{4332}\equiv(F4 - F3) - (F3 - F2) 
\end{equation}

\begin{equation}
I_{5331}\equiv(F5 - F3) - (F3 - F1) 
\end{equation}

In this document, we discuss preliminary results using the classical SDI 
data reduction method (i.e. no Fresnel propagation is considered.) 

\section{Data Acquisition and Reduction}

Data were acquired in January 2007.  A series of 5 filter images were
acquired at each nominal contrast value 
(10$^{-6}$, 10$^{-7}$, 10$^{-8}$, and 10$^{-9}$).  
High contrasts were achieved via speckle nulling in the F3 800 nm
filter.  Speckle nulling iteratively removes speckles locally by
correcting the wavefront with the deformable mirror.  Speckles are
``dialed out'' one by one by programming in the appropriate
antispeckle with the deformable mirror \citep{Mal95,Tra02,Bor06,Tra06}.  The HCIT operates at such
high contrasts that speckles due to both phase errors and amplitude 
errors in the wavefront must be corrected.  \citep[From~the~ground,~at~contrasts~of~
$\sim$10$^{-4}$ - ~10$^{-5}$,~phase~errors~dominate~so 
~completely~that~amplitude~errors~can~effectively~be~ignored, see e.g.][]{Guy05,Sou07}.  
However, only phase speckles can be dialed into the deformable mirror.
This creates a symmetry issue when removing amplitude speckles \citep{Bor06}.
In the image, both phase and amplitude 
speckles will appear as a pair of symmetric point sources around 
the central PSF.  However, when inspected in image plane electric 
field instead of image amplitude, phase and amplitude speckles 
appear rather different.  Phase speckles are antisymmetric, 
while amplitude speckles are symmetric.  Adding the opposite 
phase speckle on the DM will completely cancel out a phase speckle, but will
only cancel out one side of an amplitude speckle.  The 
other side of the amplitude speckle will be intensified \citep{Bor06}.
In our optical system, phase speckles dominate
to contrasts of 10$^{-7}$ -- up to these contrasts, a symmetric 
dark hole can be generated \citep{Tra04}.  At contrasts better than 10$^{-7}$, 
speckles are produced by a mix of phase and amplitude 
aberrations in the wavefront and the dark hole becomes asymmetric.  

\begin{figure}
   \begin{center}
   \begin{tabular}{c}
   \includegraphics[height=4cm]{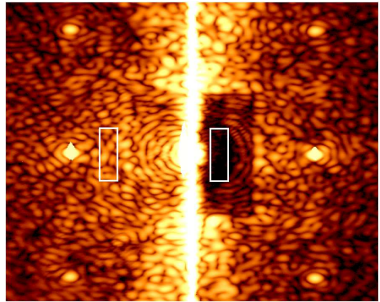} \\
   \end{tabular}
   \end{center}
   \caption[] { \label{fig:trajectories} 
	Boxes used for speckle RMS calculation (Table 2)}
   \end{figure}

\begin{figure}
   \begin{center}
   \begin{tabular}{c}
   \includegraphics[height=15cm]{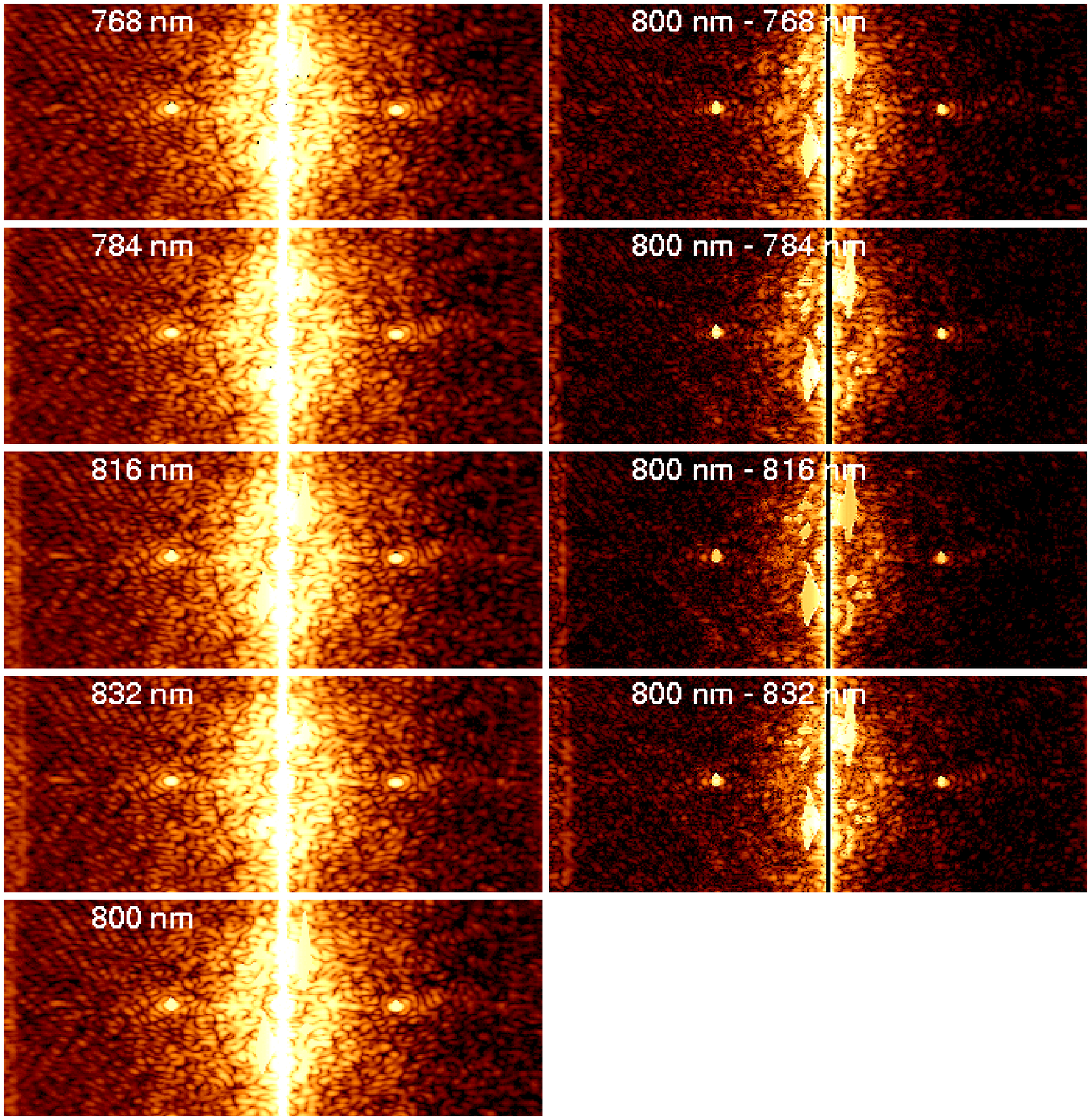}
   \end{tabular}
   \end{center}
   \caption[] { \label{fig:rogues1} Left:RAW, Right:NSDI.
	Gallery of single wavelength images and single differences
with a nominal contrast level of 10$^{-6}$.  All images
are shown with the same logarithmic stretch from contrasts of 0 to 
10$^{-5}$.  The single 
subtractions suppress the speckles outside the dark hole by a factor of
5 - 50.  The bright point sources outside the dark hole 
are ghosts produced by aliasing outside the control radius of the 
DM.
}
   \end{figure}

\begin{figure}
   \begin{center}
   \begin{tabular}{c}
   \includegraphics[height=15cm]{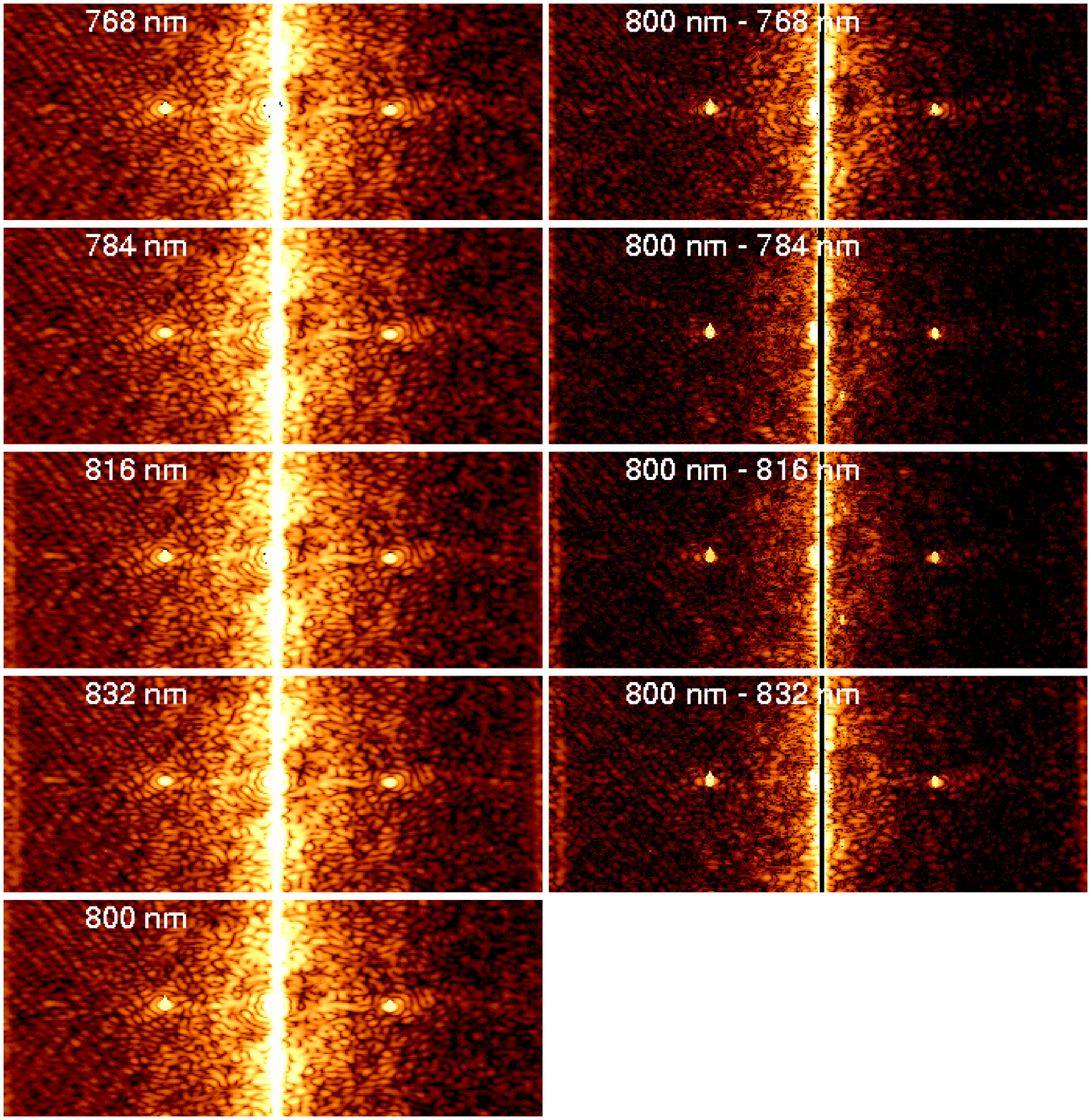}
   \end{tabular}
   \end{center}
   \caption[] { \label{fig:rogues2} Left:RAW, Right:NSDI.
Gallery of single wavelength images and single differences
with a nominal contrast level of 10$^{-7}$.  All images
are shown with the same logarithmic stretch from contrasts of 0 to 
10$^{-5}$.  The single 
subtractions suppress the speckles outside the dark hole by a factor of
5 - 50.  The bright point sources outside the dark hole 
are ghosts produced by aliasing outside the control radius of the 
DM.
}
   \end{figure}

\begin{figure}
  \begin{center}
   \begin{tabular}{c}
   \includegraphics[height=15cm]{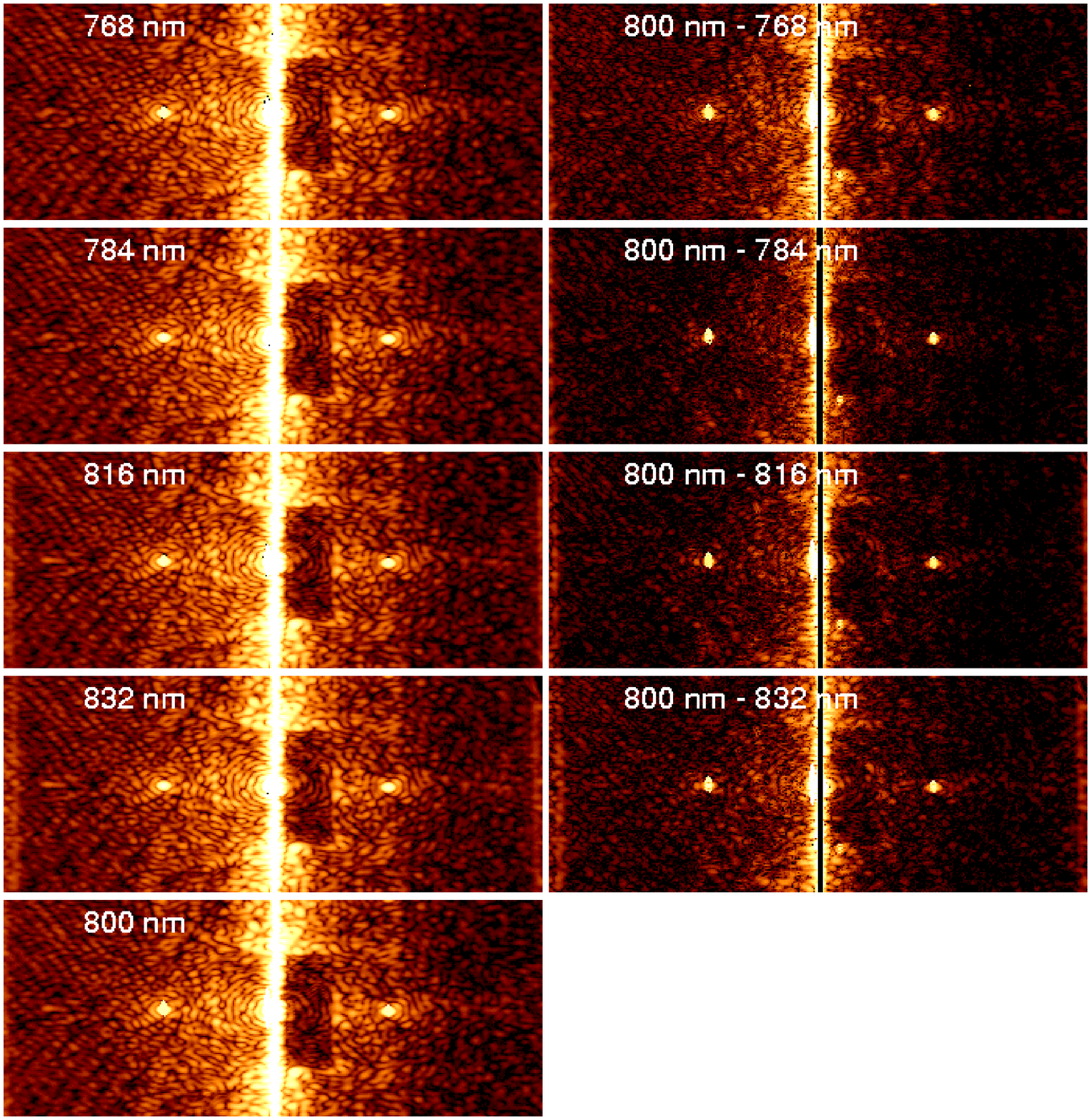}
   \end{tabular}
   \end{center}
   \caption[] { \label{fig:rogues3} Left:RAW, Right:NSDI.
Gallery of single wavelength images and single differences
with a nominal contrast level of 10$^{-8}$.  All images
are shown with the same logarithmic stretch from contrasts of 0 to 
10$^{-5}$.  The single 
subtractions suppress the speckles outside the dark hole by a factor of
5 - 50.  The bright point sources outside the dark hole 
are ghosts produced by aliasing outside the control radius of the 
DM.
}
   \end{figure}

\begin{figure}
   \begin{center}
   \begin{tabular}{c}
0   \includegraphics[height=15cm]{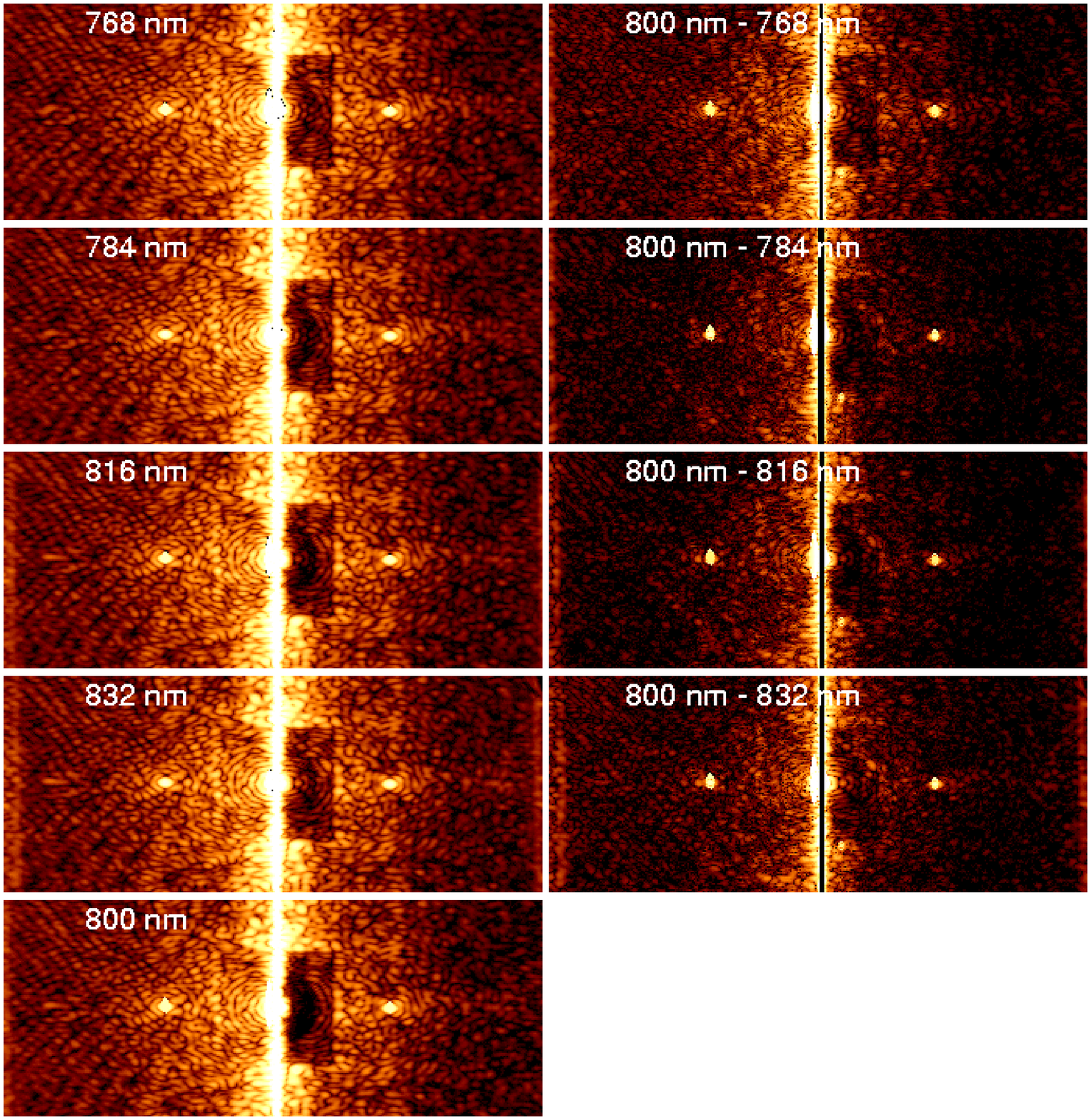}
   \end{tabular}
   \end{center}
   \caption[] { \label{fig:rogues4} Left:RAW, Right:NSDI.
Gallery of single wavelength images and single differences
with a nominal contrast level of 10$^{-9}$.  All images
are shown with the same logarithmic stretch from contrasts of 0 to 
10$^{-5}$.  The single 
subtractions suppress the speckles outside the dark hole by a factor of
5 - 50.  The bright point sources outside the dark hole 
are ghosts produced by aliasing outside the control radius of the 
DM. 
}
   \end{figure}

In this manner, the scattered light
producing the speckles can be removed locally, however, the speckle
solution is inherently asymmetric; high contrasts are achieved only in
the right-side 
dark hole region.  (Fig.~\ref{fig:trajectories} shows a schematic of 
the darkest portion of the dark hole region on the right and a comparison
region on the left.)
This is quite clear in our images (Fig.~\ref{fig:rogues1}
to Fig.~\ref{fig:rogues4}) -- at a contrast of 10$^{-6}$, a slight dark
hole is seen on both sides of the image but at contrasts of 
10$^{-7}$ or better when amplitude errors must be taken 
into account, the dark hole appears only on the right side of the image.
The size of this dark region is determined by the control 
radius of the deformable mirror, which is in turn determined by the 
total number of DM actuators and actuator spacing.

Once each contrast level was achieved via speckle nulling, two images
(with exposure times of 90 s and 5 s) were acquired at each of the five filter wavelengths. 
Per-pixel S/N ratios (for pixels representative of the average contrast value in the image dark hole)
are presented in Tab.~\ref{tab:SN}.  The gain of the CCD was 2.5 electrons/DN, 
with a read noise of 6 electrons/pix rms.
The total noise per pixel is a combination of shot noise and read noise:

\begin{equation}
N = \sqrt{N_{shot}^2 + N_{read}^2}
\end{equation}

We focus on the 90 s dataset, since its signal to noise ratio
is considerably higher (by a factor of $>$4) 
than that of the 5 s dataset.  Unfortunately, 
significant saturation within the dark hole region ($\sim$5$\%$ of pixels) 
occurred in the 10$^{-6}$ contrast images.  We display these images since 
their qualitative features are still of interest  
but exclude them from further analysis.  Some 
saturation is seen in the 10$^{-7}$ to 10$^{-9}$ contrast images, but only 
far outside the dark hole and comparison regions.  In fact, for these 
images, only ghost point sources considerably outside 
the dark hole were saturated.   (These ghost point sources are
produced by aliasing outside the control radius of the DM and are 31 $\lambda$/D
from the occulted star -- in other words, far from any analysis region).
Coronagraph images were bias and dark corrected, but no flat fielding 
was performed, as the CCD already exhibits a pixel-to-pixel DQE 
uniformity of better than 1$\%$ \citep{TT07}.  
Since the images in the nulled wavelength are of the highest quality and are
the most reliable, 
the F3 800 nm filter image was used as the ``master'' image (for adjustment 
of platescale and alignment purposes) in this analysis. 

Fluxes in each wavelength image were converted to the equivalent contrast
via the following steps: 1) a "stellar point spread function" (henceforth
PSF) was derived by 
by offsetting the coronagraph mask by 144 $\mu$m to the first maximum 
in transmission (in other words, moving the star off the mask) 
and taking a 30 $\mu$s snapshot image, 2) the "stellar
peak" was estimated 
from the snapshot image and scaled to the appropriate peak signal
for a 90 s data image, 3) each 90 s image was divided 
by the estimated peak signal 
to convert to contrast units, and 4) each peak signal corrected image was 
then divided by the analytic attenuation profile of the coronagraph mask 
(known to closely
match the measured attenuation profile) to correct for attenuation.  For 
more details on this process, see the supplemental material from \citet{TT07}.

Since the radial position of the speckle pattern in each filter image
is proportional to $\lambda$, the platescale of each image
was scaled to the 800 nm image so that the speckles in each filter
fall at the same radii despite chromatic differences.  The exact
filter central wavelengths presented in Table 1 were used for this
scaling.  To estimate the resampling noise introduced by wavelength scaling, 
we rescaled 20 randomly chosen 40$\times$60 pixel subimages
from within each of the dark holes of the 10$^{-7}$, 10$^{-8}$, and 10$^{-9}$ 
images (4 wavelengths $\times$ 3 contrast levels, so 240 subimages total 
were considered
in this analysis).  
We then calculated the standard deviation in each subimage (using robust\_sigma
in IDL) before and after scaling ($\sigma_{orig}$ and $\sigma_{scaled}$ respectively).  
Resampling noise was estimated to be:

\begin{equation}
N_{resamp} = stddev(\sigma_{orig} - \sigma_{scaled})
\end{equation}

We found values for the resampling noise of 3.5-6.6$\times10^{-11}$, which, even
at our best contrasts of 10$^{-9}$, are negligible.

\begin{figure}
   \begin{center}
   \begin{tabular}{c}
   \includegraphics[height=10cm]{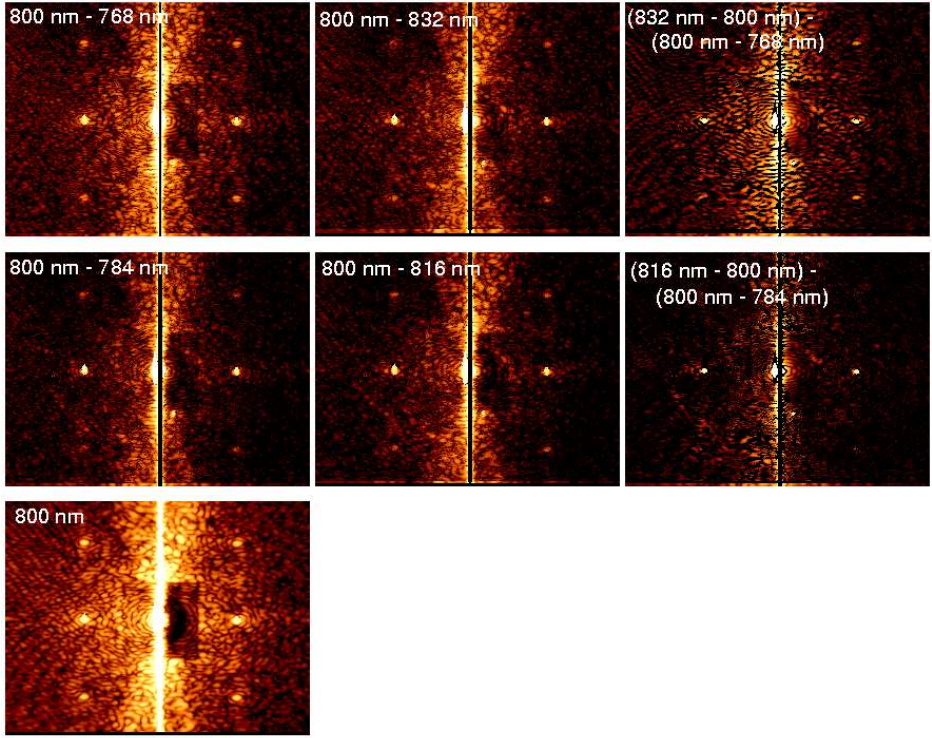}
   \end{tabular}
   \end{center}
   \caption[] { \label{fig:dd4} 
	Double differenced images -- (F4 - F3) - (F3 - F2) 
((F3 - F2) - (F2 - F1)) and (F5 - F3) - (F3 - F1) for a nominal
contrast level of 10$^{-9}$.  These images are
shown with the same logarithmic stretch from contrasts of 0
to 10$^{-5}$ as 
Fig.~\ref{fig:rogues1}.  Significant speckle residuals are present compared 
to the single differenced images.  In contrast to theory (Marois et al. 2000), 
the double difference method is less effective at 
suppressing speckle noise than a single difference.  The bright point sources outside the dark hole 
are ghosts produced by aliasing outside the control radius of the 
DM.
}
   \end{figure}

Finally, each wavelength image was aligned to the 800 nm image using a 
custom shift-and-subtract alignment algorithm (alignments are to 0.1 
pixel precision, \citet{Bil07}).  The resulting aligned images 
were subtracted from the master 800 nm image.  
Galleries of each aligned image and the resulting subtraction are
presented for each nominal contrast level in Fig.~\ref{fig:rogues1}
to Fig.~\ref{fig:rogues4}.  These images
are shown with a logarithmic stretch from contrast levels of 0 to 
10$^{-5}$.   Speckle rms measured in several 
regions outside the dark hole were reduced 
by a factor of 5-50 after subtraction (for details on speckle rms/contrast calculations, see Section 4 below).
Two double differences were also calculated and 
are shown for 10$^{-9}$ contrast in the dark hole (same logarithmic stretch, also by nominal contrast level) in Fig.~\ref{fig:dd4}.

\section{Analysis}

To quantify the level of speckle attenuation available through the 
differential imaging technique both inside and outside the dark hole, 
we calculated the contrast levels 
before and after subtraction in two
20$\times$60 pixel regions of the chip for 
all our reduced images.  Contrast (or more simply, residual speckle rms)
is estimated as the standard deviation over the region, an approach adopted from \citet{Bil07}.
One major caveat is attached to this approach -- it assumes 
Gaussian statistics, whereas speckles have been shown to follow
a Rician probability density function \citep{Aim04, Fit07}.  Using
Gaussian statistics will seriously overestimate the confidence level
of a point source detection \citep{Mar08b}.  However, this approach has
been used robustly as a diagnostic of contrast
performance \citep{Tra04,Laf07,Hin07}.  For this testbed experiment, where 
we are simply concerned with speckle residuals and not detections of actual
point sources, this method is appropriate for our use.

\begin{figure}
	\includegraphics[height=10cm]{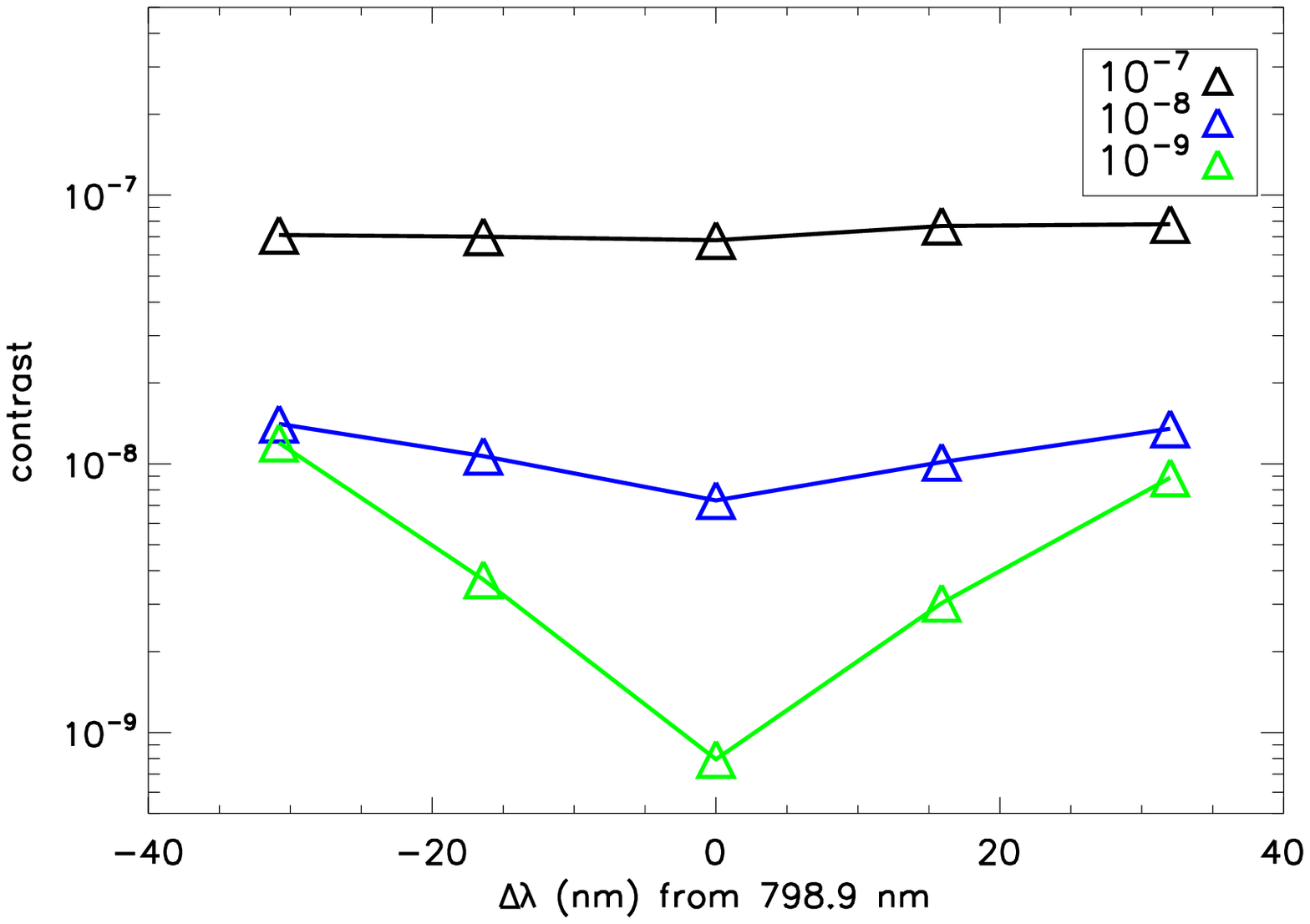}
\caption[]{ \label{fig:RMS}
	Single image 
	Contrast as a function of wavelength ($\Delta$$\lambda$ in nm from
	798.9 nm) inside the dark hole.  
	Contrast is degraded by a factor of $\geq$2 
	at wavelength differences of $\pm$30 nm from 798.9 nm.  
	The degradation of contrast
	 as a function of wavelength is most pronounced
	at the highest nulling wavelength 
	contrast level (10$^{-9}$). }
\end{figure}

Regions used for this analysis are shown in 
Fig.~\ref{fig:trajectories}.  The dark hole region lies at 4-9 $\lambda$/D
from the occulted stellar image while the left-side "comparison region"
lies 19-24 $\lambda$/D to the left of the occulted stellar image 
($\lambda$/D$=$4.67 pixels at 785 nm for the HCIT, Trauger \& Traub
2007 supplemental material).  One major caveat exists 
regarding our choice of left-side ``comparison region'' 
-- we chose a comparison region outside the control radius of the 
deformable mirror, since the speckle nulling
algorithm adds speckle noise to the left side of the image when removing 
an amplitude speckle from the right side of the image.
Before nulling
we note that a radial trend in speckle intensity exists -- there are more
and brighter speckles closer to the center of each image.  Thus, the speckle
noise within the left-side region is somewhat lower than the pre-null
speckle noise in the dark-hole region. 
For this reason, it is not instructive to 
compare speckle noise between the right and left side regions, but instead
to compare the degree of attenuation provided by the differential 
imaging technique in both of these regions.  
Contrast information is presented in 
Table 2 and is plotted for the dark hole region as a function 
of $\Delta$$\lambda$ from the nulling wavelength (800 nm) in 
Fig.~\ref{fig:RMS}. 
Outside the dark hole, the single differences 
improve contrast by a factor of $\sim$6, meaning that the NSDI method
will be highly applicable in moderate (10$^{-6}$ -- 10$^{-8}$) 
contrast systems without a DM to iteratively remove speckles 
(for instance, this technique might be very appropriate to utilize with 
a coronagraphic Jovian planet imager without a DM).
Inside the dark hole, contrast appears to depend 
strongly on $\Delta\lambda$ from
the nulling wavelength, especially at higher contrasts.  
For lower contrast levels, a similar speckle suppression with single 
subtraction
is also observed within the dark hole.
At high contrasts (10$^{-8}$,10$^{-9}$), however,
apparent chromatic variation becomes very important and the 
speckle pattern appears to decorrelate.  For instance, significantly more 
speckle noise appears in the 832 nm vs. the 800 nm image, meaning 
that a subtraction of these two images degrades the achieved 
contrast relative to the 800 nm image alone. 
For the same reason, the double differences do not provide an 
advantage over the single differences -- contrast inside and outside of 
the dark hole is not decreased compared to the single differences.  

Increased speckle noise in the dark hole as image wavelength differs from the
nulling wavelength translates to a lower contrast in that image relative
to the image at the nulled wavelength.  Our multifilter experiment
lets us simulate the variation in achieved contrast as a function 
of wavelength that we would expect within a wideband
image.  Our filters cover a wavelength range of $\sim$80 nm -- equivalent to a
10$\%$ bandwidth filter.  
At an 800 nm contrast of 10$^{-8}$, contrast is degraded by a factor of 
1.7 at the red end (830.9 nm) compared to 
the nulled wavelength (798.9 nm) and contrast is degraded by a factor of 2 
at the blue end (768.1 nm) compared to the nulled wavelength.  
Over the entire $\sim$80 nm bandwidth, contrast is on average degraded by 
a factor of 1.5 compared to contrast at the nulled wavelength 
(and, hence, the contrast would be degraded by a factor 
of $\sim$1.5$\times$ over a 10$\%$ bandwidth
filter than over a 2$\%$ bandwidth filter.)   
At the highest contrasts (10$^{-9}$) in the 800 nm image, 
contrast over the entire bandpass 
is considerably diminished 
-- at the red end (830.9 nm), contrast is degraded by a factor of 11
compared to the nulled wavelength (798.9 nm) and at the blue end (768.1 nm),
contrast is degraded by a factor of 15 compared to the nulled 
wavelength. 
Over the entire $\sim$80 nm bandwidth, contrast is degraded on average 
by a factor of 7.2 compared to the nulled wavelength. 
Thus, apparent speckle chromatism in these images
is sufficient to predict considerable contrast degradation in a wide band 
($\sim$80 nm) filter.

To determine the source of speckle chromatism in these images, we must
take into account various sources of wavefront error in the HCIT.  There
are three main categories of wavefront errors we must consider: 

{\bf 1) Wavelength scaling phase errors in the pupil plane.}  These speckles
scale according to $\lambda$ and thus can be aligned across different
wavelengths. 
Additionally, they are antisymmetric and can be completely removed by the 
deformable mirror.

{\bf 2) Wavelength scaling amplitude errors in the pupil plane.}  
These speckles scale according to $\frac{\lambda}{D}$.  They can be removed 
on one side of the image using the deformable mirror, but will 
produce an increase in the corresponding 
speckle on the other side of the image.

{\bf 3) Wavelength dependent phase errors in the focal plane due to the occulter.}  
In other words, each
wavelength sees different occulter properties.  
The current occulter is fabricated from High Energy Beam 
Sensitive (HEBS) glass, which varies in 
transmission according to wavelength and thus will introduce 
slightly different phase errors at different wavelengths.  
\citep[The~relationship~between~wavelength~and~phase~shift~for~HEBS~glass~is~plotted~in~Figure~2~of][]{Moo07} 
We hypothesize that the HEBS glass in the occulter is responsible for the
speckle chromatism observed in this experiment.
At lower contrasts 
the phase errors due to wavelength 
dependent transmission in the HEBS glass 
are negligible, but at 10$^{-9}$ contrasts, they become quite 
important.  These speckles are
thus chromatic -- they do not scale by $\lambda$
and are likely the culprit in causing the characteristic ``parabola'' 
pattern in contrast seen in Fig.~\ref{fig:RMS}. 

\begin{figure}
	\includegraphics[height=10cm]{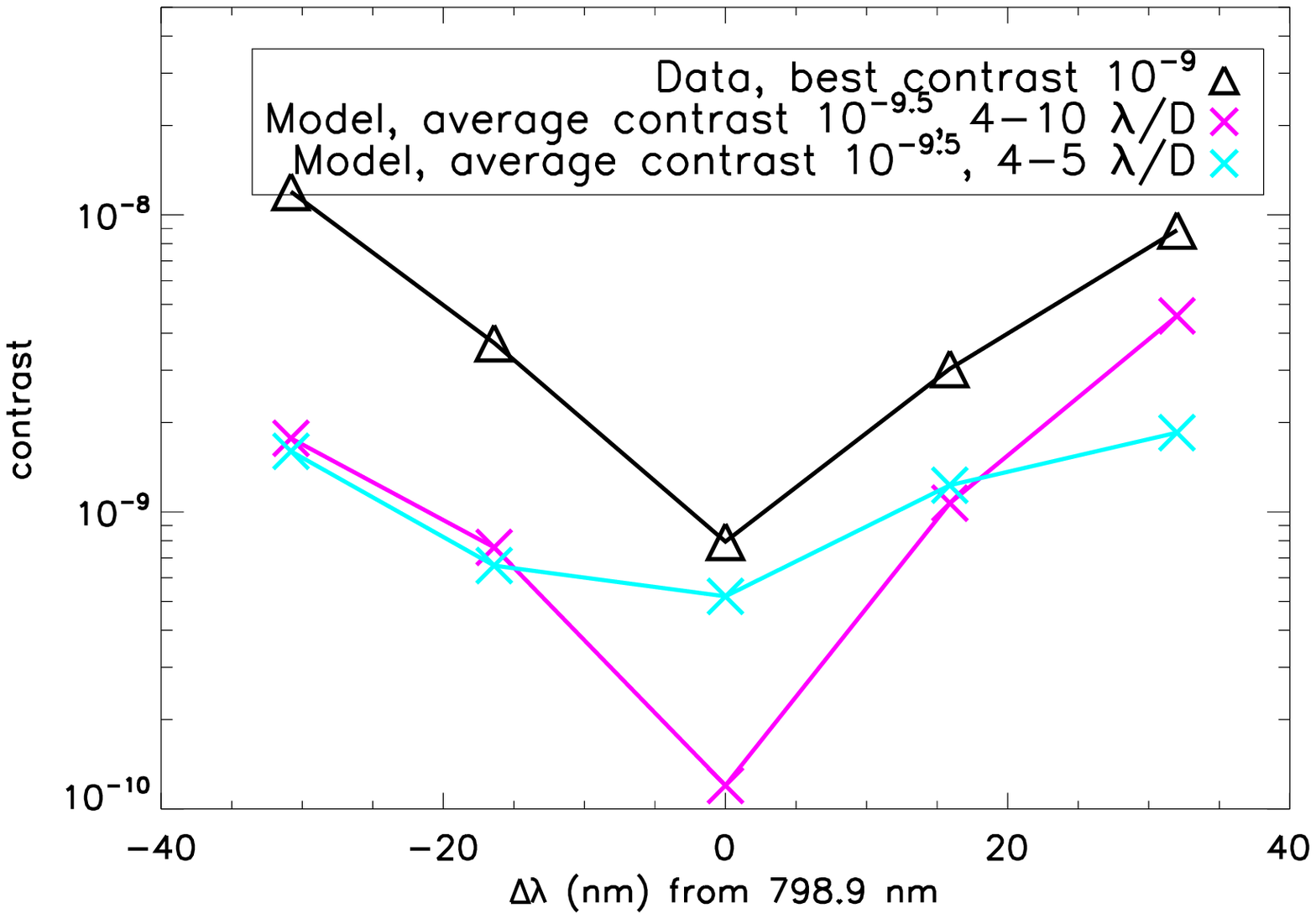}
\caption[]{ \label{fig:RMS_model}
	Single image 
	Contrast as a function of wavelength ($\Delta$$\lambda$ in nm from
	798.9 nm) inside the dark hole (10$^{-9}$ average contrast)
	and from model predictions.  
	Contrast is degraded by a factor of $\geq$2 
	at wavelength differences of $\pm$30 nm from 798.9 nm.  
	The degradation of contrast
	 as a function of wavelength is most pronounced
	at the highest nulling wavelength 
	contrast level (10$^{-9}$).  Model predictions
	for this experiment produce a qualitatively 
	similar result.  
	These models are 
	for slightly higher intrinsic contrast levels 
	calculate contrast in somewhat 
	different regions than used in the experiment, which is why there
	is some discrepancy between model and experiment results.
	The 4-5 $\lambda$/D model samples only the deepest parts of the 
	dark hole, while the 4-10 $\lambda$/D model samples a much 
	larger and somewhat less corrected region, hence, the observed
	degradation in contrast between the models.}
\end{figure}

In Fig.~\ref{fig:RMS_model},
we compare our results 
to in-house theoretical predictions for this experiment (at slightly better 
contrasts of 10$^{-9.5}$ and with somewhat different 
comparison regions) and find the same qualitative result.
The testbed optics are modeled as a series of optical surfaces, each of which has
an interferometrically measured surface map \citep[see, ~e.g.~][]{Fan05}
The occulting mask is assumed to have a profile of intensity
transmission that follows a fourth-order sinc$^2$ function \citep[][]{Kuc02}, and 
has a transmitted phase that varies as a function of intensity
transmission, as measured by \citet{Hal05}. 
Wavefront correction is performed using a model of the speckle 
nulling algorithm,
as described in \citet{Tra04}.  To simulate the image recorded on
the camera through an individual bandpass filter, a number of monochromatic
wavelengths (typically at least 3 wavelengths per filter) are propagated through
the model and summed by intensity at the camera.  The model iterates the speckle
nulling algorithm until the contrast improvements in successive iterations become
negligible.  More details regarding modeling can be found in \citet{Moo07}
and \citet{Moo08}.  Models using the current generation of 
HEBS glass occulters 
can qualitatively reproduce the speckle chromaticity 
found in the current experiment.

\citet{Moo07} used this same modeling approach to simulate 
the contrast performance of next generation hybrid coronagraph 
mask designs (metallic and 
dielectric thin films deposited on a glass substrate) using a similar
band-limited coronagraph setup and predict considerably less 
chromatic coronagraph performance from these masks.
Thus we suggest that the speckle chromaticity seen in this experiment 
is due to the $\lambda$-dependent behavior of the HEBS glass which makes 
up the occulter and not from the 
intrinsic design of the band-limited coronagraphic mask.  
\citet{Moo08} have implemented a metallic thin film mask at the HCIT (and 
achieved the best published contrasts to date at the HCIT with it).  
Hybrid masks are just becoming available at the HCIT and have already yielded
considerable improvements in contrast over a 20$\%$ bandwidth filter 
(Moody, private communication).  When 
these hybrid masks have been fully implemented at the HCIT, this 
experiment will be repeated to test this hypothesis.

\section{Discussion, Conclusions, and Future Work}

Using single wavelength speckle nulling with a band-limited Lyot
coronagraph and operating 
at nominal contrasts of 10$^{-6}$ to 10$^{-9}$, single differences of
filter images can reduce speckle noise outside of the dark hole by
factors of 5 - 50.  For contrasts of 10$^{-6}$ to 10$^{-8}$, a similar
result is also found within the dark hole, with speckle attenuation
achieved through a single difference decreasing as a function of
increasing contrast.  However, at high contrasts (10$^{-9}$),
considerable increase in speckle RMS between filters is observed to
``pollute'' the dark hole in all of our single differences (where, in
each difference, only one of the two wavelengths had undergone optical
speckle nulling).  At all contrast levels, a double difference of
images does not seem to decrease speckle noise relative to the single
differences.  Significant differences  
in contrast ($\sim$2-10$\times$RMS) are
found between filters separated in wavelength by $\Delta$$\lambda$
$>$20 nm.  Contrast degradation
 between filters increases strongly as a function of 
increasing contrast at the nulling wavelength.  
This chromatic speckle noise is likely due to the chromaticity of the 
materials of the occulter itself.

However, this experiment is only one possible implementation using
specific choices of coronagraphic mask/occulter and nulling algorithms
available in January 2007.  It is clear that the occulter used in
January 2007 possessed significant wavelength dependent phase errors,
causing speckle decorrelation between filters.  This decorrelation was
exacerbated by using a speckle nulling algorithm which calculated a
speckle solution at only one wavelength.  Since 2007, significant
effort has gone into developing non-chromatic coronagraphic
focal plane occulters and nulling algorithms for the HCIT.  Thus, another
implementation of this experiment using these improvements would
likely be able to extend NSDI performance to even better contrasts.
Specific improvements possible include:

1) The major limiting factor for this experiment is likely
chromatism due to the materials of the 
occulter itself.  However, since January 2007, 
a variety of less chromatic occulter options are under development.
\citet{Moo07} explore hybrid coronagraph mask designs (metallic and 
dielectric thin films deposited on a glass substrate)
which promise considerably less chromatic coronagraph performance.
These masks are still in development but will in principle allow the extension
of differential imaging techniques to 10$^{-9}$ -- 10$^{-10}$ contrasts.

2) Algorithms which produce a multiwavelength speckle solution are now
available \citep[e.g.~multiwavelength~electric~field~conjugation,][]{Giv07}
instead of speckle nulling at a specific wavelength.  Utilizing
one of these newer algorithms will produce a 
broadband solution without degradation at either end of the 
bandpass.

A future iteration of this experiment will be conducted once these
improvements are implemented.

\acknowledgements

Part of the research described in this paper was carried out at the 
Jet Propulsion Laboratory, California Institute of Technology, under 
a contract with the National Aeronautics and Space Administration.
Support for B.B. was provided by NASA 
through the NASA GSRP grant NNG04GN95H and the Hubble Fellowship grant 
HST-HF-01204.01-A awarded by the Space Telescope Science Institute, 
which is operated 
by the Association of Universities for Research in Astronomy, Inc., for NASA, 
under contract NAS 5-26555.

\clearpage

\begin{deluxetable}{lcc}
\tablecolumns{3}
\tablewidth{0pc}
\tablecaption{Filter Wavelengths and Bandwidths}
\tablehead{
\colhead{Filter} & \colhead{Center Wavelength} & \colhead{FWHM}}
\startdata
F1 & 768.1 nm & 15.1 nm \\ 
F2 & 782.5 nm & 15.6 nm \\      
F3 & 798.9 nm & 15.9 nm \\
F4 & 814.8 nm & 16.3 nm \\
F5 & 830.9 nm & 15.5 nm \\ \hline
\enddata
  \label{tab:filtprop}
\end{deluxetable}

\begin{deluxetable}{lcccccc}
\tablecolumns{6}
\tablewidth{0pc}
\tablecaption{Per-Pixel S/N Ratio in the Dark Hole}
\tablehead{
\colhead{Contrast} & \colhead{Exp. Time} & \colhead{F1} & \colhead{F2} & \colhead{F3} & \colhead{F4} & \colhead{F5}}
\startdata
10$^{-6}$ pixel in 10$^{-6}$ image & 90 s & 179.9 & 182.4 & 179.9 &  187.9 & 142.2 \\
10$^{-7}$ pixel in 10$^{-7}$ image & 90 s &  45.6 &  47.1 & 45.8 & 68.3 & 68.8 \\
10$^{-8}$ pixel in 10$^{-8}$ image & 90 s &  21.2 &  21.5 & 19.2 & 19.2 & 18.1 \\
10$^{-9}$ pixel in 10$^{-9}$ image & 90 s &   4.4 &   2.5 &  2.7 &  2.5 &  2.7 \\\hline
10$^{-6}$ pixel in 10$^{-6}$ image &  5 s &  42.0 &  42.6 & 42.0 & 43.9 & 33.0 \\
10$^{-7}$ pixel in 10$^{-7}$ image &  5 s &   9.5 &  9.9  &  9.5 &  15.2 & 15.3 \\
10$^{-8}$ pixel in 10$^{-8}$ image &  5 s &   3.4 &  3.5  & 2.9 &  2.9 & 2.7 \\
10$^{-9}$ pixel in 10$^{-9}$ image &  5 s &   0.3 &  0.2  & 0.2 &   0.2 & 0.2 \\\hline
\enddata
  \label{tab:SN}
\end{deluxetable}

\begin{deluxetable}{ccccc}
\tablecolumns{6}
\tablewidth{0pc}
\tablecaption{Speckle RMS in Right-Side Dark Hole (dh) and Left-Side Comparison Region (oh)}
  \label{tab:specproperties6}
\tablehead{
\colhead{Image} & \colhead{Region} & \colhead{10$^{-7}$} & \colhead{10$^{-8}$} & \colhead{10$^{-9}$}}
\startdata
F1 768.1 nm & dh  & 7.1$\times 10^{-8}$ & 1.4$\times 10^{-8}$ & 1.2$\times 10^{-8}$ \\
 & oh & 4.0$\times 10^{-8}$ & 4.7$\times 10^{-8}$ & 5.4$\times 10^{-8}$ \\ \hline
F2 782.5 nm & dh & 7.0$\times 10^{-8}$ & 1.1$\times 10^{-8}$ & 3.7$\times 10^{-9}$
\\  
 & oh & 3.7$\times 10^{-8}$ & 4.3$\times 10^{-8}$ & 5.0$\times 10^{-8}$ \\ \hline    
F3 798.9 nm (nulled $\lambda$) & dh & 6.9$\times 10^{-8}$ & 7.3$\times 10^{-9}$ & 7.9$\times 10^{-10}$\\
 & oh & 3.7$\times 10^{-8}$ & 4.4$\times 10^{-8}$ & 4.9$\times 10^{-8}$ \\ \hline
F4 814.8 nm & dh & 7.7$\times 10^{-8}$ & 1.0$\times 10^{-8}$ & 3.0$\times 10^{-9}$ \\
 & oh & 3.6$\times 10^{-8}$ & 4.2$\times 10^{-8}$ & 4.7$\times 10^{-8}$\\ \hline
F5 830.9 nm & dh & 7.8$\times 10^{-8}$ & 1.3$\times 10^{-8}$ & 8.9$\times 10^{-9}$ \\
 & oh & 3.5$\times 10^{-8}$ & 4.1$\times 10^{-8}$ & 4.7$\times 10^{-8}$ \\ \hline
798.9 nm - 768.1 nm & dh & 1.7$\times 10^{-8}$ & 1.0$\times 10^{-8}$ & 1.2$\times 10^{-8}$ \\
 & oh & 6.4$\times 10^{-9}$ & 6.8$\times 10^{-9}$ & 9.3$\times 10^{-9}$ \\ \hline
798.9 nm - 782.5 nm & dh & 7.5$\times 10^{-9}$ & 3.9$\times 10^{-9}$ & 3.0$\times 10^{-9}$ \\
 & oh & 2.6$\times 10^{-9}$ & 2.9$\times 10^{-9}$ & 3.5$\times 10^{-9}$ \\ \hline
798.9 nm - 814.8 nm & dh & 7.6$\times 10^{-9}$ & 3.8$\times 10^{-9}$ & 2.5$\times 10^{-9}$ \\
 & oh & 2.7$\times 10^{-9}$ & 3.4$\times 10^{-9}$ & 4.0$\times 10^{-9}$ \\ \hline
798.9 nm - 830.9 nm & dh & 1.3$\times 10^{-8}$ & 8.6$\times 10^{-9}$ & 8.5$\times 10^{-9}$ \\
 & oh & 4.7$\times 10^{-9}$ & 5.3$\times 10^{-9}$ & 5.8$\times 10^{-9}$ \\ \hline
(814.8 nm - 798.9 nm) & dh & 8.0$\times 10^{-9}$ & 3.7$\times 10^{-9}$ & 4.8$\times 10^{-9}$ \\ 
 - (798.9 nm - 782.5 nm) & oh & 5.2$\times 10^{-9}$ & 5.2$\times 10^{-9}$ & 5.4$\times 10^{-9}$ \\ \hline
(830.9 nm - 798.9 nm)  & dh & 1.8$\times 10^{-8}$ & 1.3$\times 10^{-8}$ & 1.9$\times 10^{-8}$ \\
 - (798.9 nm - 768.1 nm) & oh & 9.0$\times 10^{-9}$ & 1.1$\times 10^{-8}$ & 1.4$\times 10^{-8}$ \\ \hline
 \enddata
\end{deluxetable}

\clearpage

\clearpage

\clearpage

\end{document}